\documentclass [12pt]{iopart}
\usepackage{graphicx}

\begin{document}
\title[On the average charge of the oxygen vacancy in perovskites]
{On the average charge of the oxygen vacancy in perovskites
necessary for kinetic calculations}

\author{ S. A. Prosandeev}

\address{Physics Department, Rostov State University, 344090 Rostov
on Don, Russia.\\ Ceramics Division, National Institute of
Standards and Technology, Gaithersburg, MD}

\begin{abstract}
An analytical result has been obtained for the value of the
dynamical charge necessary for calculations of oxygen vacancy
kinetics in dielectric perovskite-type crystals. It is shown by
using the Berry phase analysis that this charge equals the nominal
charge of the vacancy: for example, for the double charged state,
it is 2; a neutral vacancy has the zero charge.
\end{abstract}

\maketitle

Oxygen vacancies in oxides of the perovskite family can be easily
created or removed by thermal treatment in   reduction or
oxidizing atmosphere respectively \cite{1}. The vacancies are
always present in single crystals and even more so in ceramics and
thin films. This nonstochiometry effects some properties of
perovskite-type oxides: dielectric permittivity can have a
noticeable \cite{2,3,4} and sometimes even large \cite{5}
contribution from electrons connected with the oxygen vacancies;
Second Harmonic Generation and luminescence experiments revealed a
large effect of oxygen vacancies on the intensities \cite{6,7};
Electroconductivity of the samples is strongly dependent on the
oxygen vacancy concentration an, hence, on thermal treatment
\cite{8}.

Sometimes the presence of oxygen vacancies helps to improve
properties of materials, for instance, it is possible to obtain
extremely high dielectric permittivity in a wide temperature
interval due to Maxwell-Wagner relaxation \cite{5}. It happens
that these unavoidable point defects spoil  some properties, for
example, in ferroelectric memory applications the presence of
oxygen vacancies is thought to be crucial and results in the
phenomenon called ``fatigue'' \cite{9,10}: ferroelectric thin
films degrade after a large number of polarization switching due
to the movement of the oxygen vacancies leading to the creation of
large frozen internal fields preventing the switching process.

In order to study the last effect one can use kinetic equations
describing a charge transport in a bias field \cite{11}. Charged
oxygen vacancies are thought to be moved in this field
\cite{9,10}. For quantitative estimations one needs to know the
oxygen vacancy charge $Z$ and the field $E_{a}$ acting on the
vacancy. In review \cite{12} this problem was discussed in
details: there had been several trials to describe these
quantities among which there are two very different proposals, one
of them suggests using Lorentz's expression for the local field,
$E_a = (\varepsilon + 2)E / 3$, and the other assumes that each of
the vacancies can be surrounded by a sphere and the field inside
this sphere is described by an Onsager's expression: $E_a =
3\varepsilon / (2\varepsilon + 1)E$~ where $\varepsilon $~ is
dielectric permittivity and $E $~ being the bias macroscopic
field. Scott and Dawber \cite{12} criticized the first approach as
giving an unrealistic diverging value of the local field at large
$\varepsilon $~ inherent to ferroelectrics, and they suggested
using the Onsager expression which much better suits experimental
data.

In this letter I will show that the problem of choosing the local
field and charge of the vacancy can be solved analytically with
the help of a Berry phase analysis \cite{13}. This approach
rigorously considers not only the local field effect but also the
covalent effect connected with the covalent binding of the ions.
The resultant polarization is computed quantum mechanically by
taking integrals over the Brillouin zone.

Instead of dealing with local fields it is easier to consider
average, macroscopic, field. In this case the charge of the
vacancy must be replaced by the dynamical charge

\begin{equation}
\label{eq1}
Z_{i\alpha ,\beta }^\ast = \frac{\partial P_\alpha }{\partial r_{i\beta } }
\end{equation}

\noindent where $P_\alpha $ is $\alpha $-th component of
polarization, and $r_{j\beta } $ is the displacement of $i$-th ion
in the $\beta $-th direction. As the average field is uniform,
only the dynamical charge should be averaged over a path of the
oxygen vacancy

\begin{equation}
\label{eq2}
Z = \frac{1}{l}\int {Z_{V_O zz}^\ast ({\rm {\bf r}})dr}
\end{equation}

\noindent
where $l$ is the length of this path; for the sake of simplicity, we directed
the field along the $z$ axis. By substituting definition (\ref{eq1}) to the integral in
(\ref{eq2}) one has

\begin{equation}
\label{eq3}
Z = \frac{1}{l}\int {\frac{\partial P_z }{\partial r_{V_O z} }dr =
\frac{\Delta P_z }{l_z }}
\end{equation}

\noindent
where $\Delta P$ is the finite difference in polarization after performing
the run over the path, and $l_{z}$ is the z component of the displacement.

The difference in polarization $\Delta P$ is the sum of
polarization stemming from the ionic charges transferred, $\Delta
P_i = Z_i l_z $~ where $Z_i$~ is the nominal (ionic) charge of the
$i$-th particle, and the electronic polarization $\Delta P_e $:

\begin{equation}
\label{eq4}
\Delta P_z = \Delta P_i + \Delta P_e
\end{equation}

The computation of the electronic (covalent) contribution can be
carried out with the help of the Berry phase analysis \cite{13}:

\begin{equation}
\label{eq5}
\Delta P_e \sim \Delta \varphi
\end{equation}

\noindent
where $\varphi $ is the average over the electronic bands Berry phase. The
Berry phase strongly changes inside the unit crystal cell and it is a
periodic function with respect to the displacement with the period
coinciding with the lattice parameter.

Consider the path shown in Fig. 1. The initial and final states in this path
have the same point symmetry and, consequently, the electronic contribution
vanishes in this path: $\Delta P_e = 0$. In other words, at the
displacements by a lattice constant or by a few lattice constants, the
electronic contribution to the polarization vanishes and only the ionic
contribution remains. This rigorously proves that the average dynamical
charge of the oxygen vacancy equals its nominal charge; for example, for the
double charged vacancy $V_O^{''} $, this charge is 2, for the single charged
vacancy it equals 1, and, finally, for the neutral vacancy it is zero.

It follows from the analyses performed that the Lorentz expression
for the local field provides wrong result for the kinetics of
oxygen vacancies in oxides. This expression gives a very large
value which, in fact, reflects the fact of strong local field in
the cenrosymmetric position of the simple cubic lattice made of
polarizable ions. If one averages this field over the unit cell,
as it was evenly suggested in Ref. \cite{12} then the difference
between the local field and the average one obviously vanishes at
all. Notice that we have shown that it is enough even to average
the result over any of the paths coinciding two points in the
lattice with equal point symmetry. Hence the field averaged over
such a path equals the field averaged over the unit cell volume.
Close result can be obtained if one uses the Onsager expression at
high $\varepsilon $: this result is only 3/2 times larger with
respect to the right expression which uses the average field
instead of the local one.

I appreciate RFFB {grant \#16021} for partial support and Umesh
Waghmare for discussions.

\section*{Caption}

Fig. 1. A path of the oxygen vacancy between equal points in the
perovskite-type lattice. Large circles correspond to B ions in the
ABO$_3$~ perovskite structure, small ones to oxygens, and the
square is the oxygen vacancy.


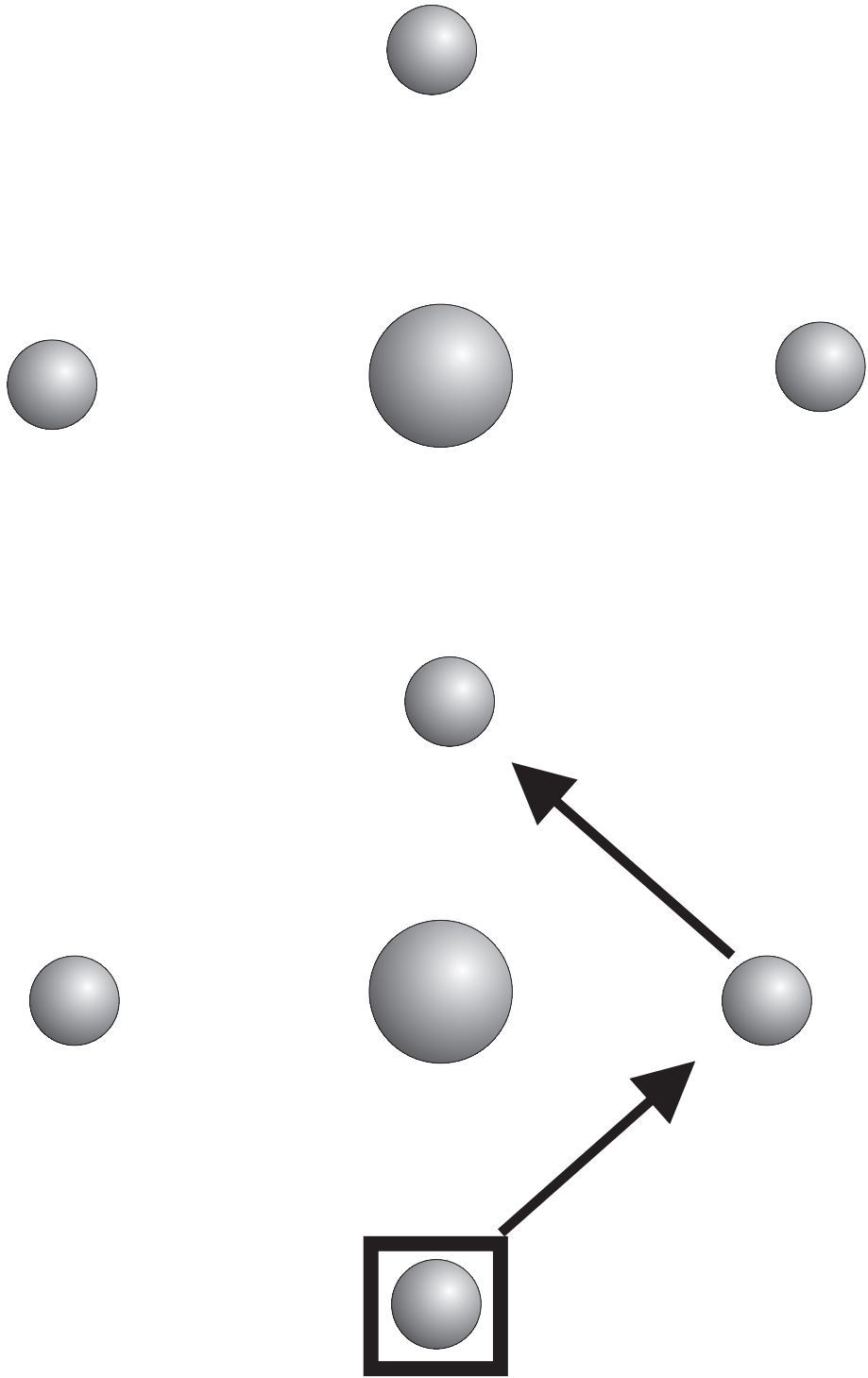
\end{document}